\documentclass{ws-procs10x7}
\usepackage{balance}

\newcolumntype{d}[1]{D{.}{.}{#1}}

\newcommand{\be}{\begin{equation}}
\newcommand{\ee}{\end{equation}}
\newcommand{\bea}{\begin{eqnarray}}
\newcommand{\eea}{\end{eqnarray}}
\newcommand{\beas}{\begin{eqnarray*}}
\newcommand{\eeas}{\end{eqnarray*}}
\newcommand{\lamqcd}{\Lambda_\mathrm{QCD}}
\newcommand{\ti}{{T^{\rm I}_i}}
\newcommand{\tii}{{T^{\rm II}_i}}
\newcommand{\eq}[1]{{(\ref{#1})}}

\makeindex
\begin{document}

\title{PENGUIN AMPLITUDES IN HADRONIC $B$ DECAYS:\\
NLO SPECTATOR SCATTERING}

\author{SEBASTIAN J{\"A}GER$^*$
}

\address{Institut f\"ur Theoretische Physik E, RWTH Aachen,
D-52056 Aachen, Germany\\
{\em Address after 01 October 2006:}
Arnold Sommerfeld Center, Department f\"ur Physik,
Ludwig-Maximilians-Universit\"at M\"unchen, Theresienstra{\ss}e 37,
D-80333 M\"unchen, Germany\\
$^*$E-mail: jaeger@theorie.physik.uni-muenchen.de}


\twocolumn[\maketitle\abstract{We present results on the
NLO ($\alpha_s^2$) spectator-scattering
corrections to the topological penguin amplitudes for charmless
hadronic two-body $B$-decays in QCD factorization.
The corrections can be sizable for the
colour-suppressed electroweak penguin amplitudes $\alpha_{4, \rm EW}^p$
but otherwise are numerically small. Our results explicitly
demonstrate factorization at this order. To assess the
phenomenological viability of the framework, we consider
penguin-to-tree
ratios in the penguin-dominated $\pi K$ system and find agreement to the
expected precision (i.e., a power correction).
}
]

\section{Introduction}
Branching ratios and CP asymmetries in $B$-decays into two
light mesons\footnote{We restrict ourselves to
  flavour-$SU(3)$-nonsinglet mesons in this note.}
provide access to the flavour structure of the Standard Model
(CKM matrix elements) and its possible extensions. Within the Standard
Model, the theoretical expressions always involve two terms with a
relative weak phase,
\be
   {\cal A}(\bar B \to M_1 M_2) = T_{M_1 M_2} e^{-i \gamma} + P_{M_1
     M_2} .
\ee
Direct CP asymmetries are then governed, besides $\gamma$,
by the imaginary part ${\rm Im}\, P/T$.
A nontrivial theoretical task is to evaluate the strong amplitudes
$P$ and $T$. Integrating out the weak scale by means of the weak
effective Hamiltonian, this reduces to the computation of hadronic matrix
elements of local operators $Q_i$, a task currently not feasible on the
lattice. Fortunately,
at leading power in an expansion in $\lamqcd/m_b$ they obey\cite{BBNS}
\bea				\label{eq:factform}
\lefteqn{  \langle M_1 M_2 | Q_i | \bar B \rangle}
 \\
	&=& F^{B  M_1}(0) \!\! \int \! {\rm d}u\, \ti(u)\, \phi_{M_2}(u)
		\; + (M_1 \leftrightarrow M_2)
\nonumber \\
	&+& \! \int \! {\rm d} \omega\, {\rm d}u\, {\rm d} v
		\tii(\omega, v, u) \phi_{B}(\omega) \phi_{M_1}(v)
                \phi_{M_2}(u) . \nonumber
\eea
The hard-scattering kernels $\ti,\,\tii$ are perturbatively calculable
as series in $\alpha_s$, while the form factors $F^{B  M_{1}}$ and the
light-cone distribution amplitudes (LCDAs) $\phi$ encapsulate universal
nonperturbative
properties of the initial- and final-state particles. An important
outcome is that
all strong phases are contained in the hard kernels.
$\ti$ is
currently known to ${\cal O}(\alpha_s)$,\cite{BBNS} while the computation
of $\tii$\cite{BBNS,BJ05,Kivel:2006xc,BJ06}
has recently been completed
at ${\cal O}(\alpha_s^2)$ by
evaluating the one-loop spectator-scattering corrections to the
(topological) penguin amplitudes.\cite{BJ06} As stated above,
the latter are crucial for any direct CP asymmetry in the Standard
Model. Moreover they are important for the branching fractions
particularly of the penguin-dominated $\Delta S=1$ modes such as $\bar
B\to \pi \bar K$. Spectator-scattering
contributions to their strong phases and effects proportional to the large
Wilson coefficient $C_1$ appear first at this order, similarly to
the colour-suppressed tree amplitude considered in Ref.~\refcite{BJ05}.

\section{Effective theory and matching}
The kernels $\tii$ receive contributions from two
hard scales $\mu_b \sim m_b$, $\mu_{\rm hc} \sim \sqrt{m_b
  \Lambda}$. They correspond to hard and hard-collinear terms in an
expansion by momentum regions.\cite{Beneke:1997zp} The result is
further factorization
$\tii(\omega, v, u) = \int {\rm d}z H_i^{\rm II}(\mu_b; u, z) J(\mu_{\rm hc};
\omega, v, z)$.\cite{Chay:2003ju,Bauer:2004tj}
The coefficient functions are conveniently obtained in a two-step
matching onto soft-collinear effective theory,\cite{scetmomspace,scetposspace}
QCD $\to$ SCET$_{\rm I}$ $\to$ SCET$_{\rm II}$, integrating out
subsequently the hard and hard-collinear scales. The hard coefficients
$\ti$ and $H_i^{\rm II}$ are interpreted as Wilson coefficients in
SCET$_{\rm I}$ and are found by solving the matching equation
\begin{eqnarray}      \label{eq:match}
  Q_i &=& \int\! {\rm d}t\, T^{\rm I}_i(t) [\bar \chi(t n_-) \chi(0)]
     \Big[C_{A0}\, [\bar \xi(0) h_v(0)]
 \nonumber \\
&& \!\!\!\!\! -\frac{1}{m_b} \int\!{\rm d}s\, C_{B1}(s) [\bar \xi(0)
  \,D_{\perp \rm hc1}(s n_+) h_v(0)]\Big]
\nonumber \\
&& \!\!\!\!\!\!\!\!  + \frac{1}{m_b} \int\! {\rm d}t {\rm d}s\,
      H^{\rm II}_i(t,s)\, [\bar \chi(t n_-) \chi(0)] \nonumber \\
&& \qquad \qquad \times [\bar \xi(0)
  \,D_{\perp \rm hc1}(s n_+) h_v(0)] ,
\end{eqnarray}
where the (schematic) rhs involves SCET$_{\rm I}$ collinear
fields for the directions of
motion of $M_2$ ($\chi, \bar \chi$) and  $M_1$ ($\xi, A_{\perp \rm
  hc1}$) as well as soft fields $h_v$ and $\bar q_s$ suitable for
interpolating the $B$-meson,
sandwiched between suitable partonic states. 
The peculiar form
of the second bracket in the first convolution is designed to
reproduce full-QCD heavy-to-light form factors as in~\eq{eq:factform}.
Of interest to us is the second term in~\eq{eq:match}, which includes
all hard interactions of $M_2$ with the spectator quark.
Decoupling properties of
SCET$_{\rm I}$ suggest that its hadronic matrix element factorizes
into a light-cone distribution amplitude $\langle M_2 | [\bar \chi
\chi] | 0 \rangle \propto \phi_{M_2}$ and a nonlocal object
$\Xi^{B M_1}(s) = \langle M_1 | [\bar \xi(0)
  \,D_{\perp \rm hc1}(s n_+) h_v(0)] | \bar B \rangle$.
This expectation is indeed confirmed by the finiteness of the
convolutions, found in all currently available computations.

Several fermion-line topologies occur in the full-QCD amplitude
differing in how the quark fields in $Q_i$ are contracted with the
fields interpolating the external states and/or with each other.
Each pair ($Q_i$, topology) contributes to one of
a few operators in the effective theory that are distinguished,
besides the chirality of the light fields, only by
their flavour content. Their matrix elements define scale- and
scheme-independent amplitude coefficents $\alpha_i$.
One-loop spectator-scattering corrections to the colour-allowed and
colour-suppressed topological ``trees'' $\alpha_{1}$, $\alpha_2$ have been
computed in Refs.~\refcite{BJ05,Kivel:2006xc},  and
the corresponding
QCD penguin amplitudes $\alpha_3^p$, $\alpha_4^p$ ($p=u,c$)
as well as the electroweak
penguin amplitudes $\alpha_{3, \rm EW}^p$, $\alpha_{4, \rm EW}^p$ are given
in Ref.~\refcite{BJ06}.

To arrive at the final form~\eq{eq:factform} one has to perform a
second matching step\footnote{Alternatively, one could try to extract
$\Xi^{B M_1}(s)$ from experiment.\cite{Bauer:2004tj}
This is not feasible beyond LO because the full $s$-dependence is needed.}
corresponding to the matching equation
\bea \nonumber 
\lefteqn{ \int\!{\rm d}^4 x\,  T\Bigg({\cal L}^{(1)}_\mathrm{SCET_I}(x)
      [\bar \xi(0) \,D_{\perp \rm hc1}(s n_+) h_v(0)] \Bigg) }
 \\
&=& \!\!
   \int \!\! {\rm d}w {\rm d}r J(w, r)  [\bar \xi(r n_+) \xi(0)] [\bar q_s(w n_-) h_v(0)].
\nonumber
\eea
The jet function $J$ containing the hard-collinear physics appears identically
in the factorization formula for form factors\cite{Beneke:2003pa}
and is known to NLO\cite{jet},
while the brackets $[\bar \xi \xi]$ and $[\bar q_s h_v]$ result in
LCDAs for $M_1$ and the $B$-meson once hadronic matrix elements are taken.

\section{Penguin amplitudes}
\begin{figure}[t]
\centerline{\psfig{file=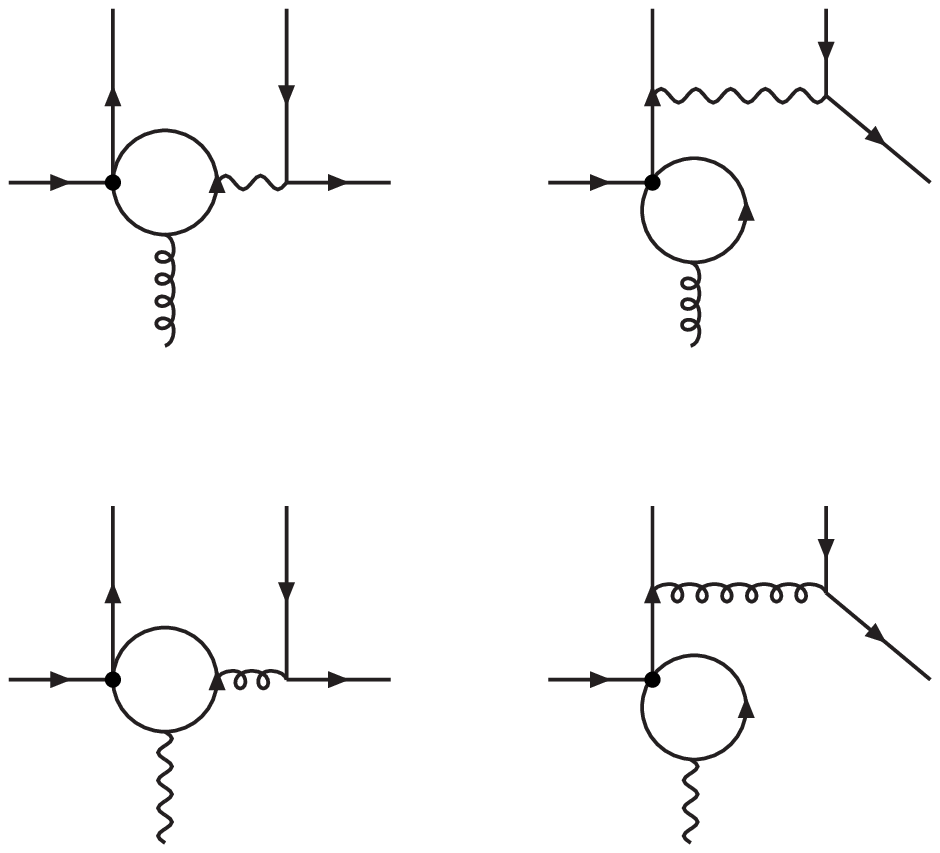,width=5.0in}}
\caption{Sample diagrams related to the penguin amplitudes.
The upper row contributes to $\alpha^p_{4,\rm EW}$ while the lower row
introduces isospin breaking in $\alpha_4^p$.}
\label{fig1}
\end{figure}
Some diagrams related to computing the coefficients $H_i^{\rm II}$
relevant to the penguin amplitudes are shown in Fig.~\ref{fig1}. In
each diagram, the
line to the left is the $b$-quark, the upgoing lines are collinear
with $M_2$, the right-going line (hard-)collinear with $M_1$, and
the external vector boson is a hard-collinear-1 gluon or photon.
Of the first row in the Figure, the left diagram contributes to the
(colour-suppressed) electroweak penguin amplitude $\alpha_{4,\rm
  EW}^p$ ($p=u$ or $c$), due to the photon exchanged between
quark lines.
Tadpole diagrams like the one on the right vanish when summed.
On the other hand, the
diagrams on the second row induce the flavour structure of the
QCD penguin amplitude $\alpha_4^p$ due to the exchanged gluon,
but the hard-collinear
photon implies isospin breaking. Such small electromagnetic corrections
to the QCD penguin amplitudes are omitted here.\footnote{Including them
consistently would necessitate taking into account isospin-breaking
effects in the form factors, LCDAs, and decay constants, among other
complications.}
When the photons in the upper row are replaced by gluons,
contributions to $\alpha_4^p$ arise. There are more diagrams,
including ones without quark loops but with
insertions of penguin operators from the weak Hamiltonian.
With the input parameter ranges given in Ref.~\refcite{BJ06} we obtain for the
leading-power contribution $a_4^c$ to the amplitude
$\alpha_4^c$:
\bea
 \lefteqn{ a_4^c(\pi \pi) = -0.029} \nonumber \\
 && - [0.002 + 0.001i]_V - [0.001+0.007i]_P
\nonumber \\
&&   +\left[ \frac{r_{\rm sp}}{0.485} \right] \!\!
   \Big\{ [0.001]_{\rm LO} + [0.001 +0.001i]_{HV} \nonumber \\
&& \qquad \quad + [0.000-0.000i]_{HP}
   + [0.001]_{\rm tw3} \Big\}
\nonumber \\
&=& -0.028^{+0.005}_{-0.003} + (-0.006^{+0.003}_{-0.002})i .
\eea
The contributions labeled ``$V$'' and ``$P$'' originate from one-loop
vertex and penguin corrections to $\ti$ in the first (form-factor) term
in~\eq{eq:factform}, while the terms ``$HV$'' and ``$HP$'' denote the
newly computed one-loop spectator-scattering corrections. The term
``tw3'' denotes an estimate of the twist-3 tree-level spectator
scattering contribution, which while being a power correction
is by convention included $a_4^p$.
The numbers show that the impact of the new corrections is very small.
This is somewhat surprising as the large Wilson coefficient $C_1$ is
involved in the ``$HP$'' terms. Closer inspection shows a numerical cancellation
between diagrams carrying different colour factors, the origin of which
is unclear.
The corrections to the colour-allowed electroweak penguin
amplitude $\alpha_{3, \rm EW}$ are also very small. The colour-suppressed electroweak penguin
amplitude $\alpha_{4, \rm EW}$ receives a larger correction. The
correction to its leading-power part $a_{10}$
can be ${\cal O}(100 \%)$ with respect to the ${\cal O}(\alpha_s)$
result.\cite{BJ06}
This is because, like the colour-suppressed tree
amplitude $\alpha_2$, $\alpha_{4, \rm EW}$ is especially
sensitive to spectator scattering due to a numerical cancellation
between naive factorization and the 1-loop correction to the first
term in~\eq{eq:factform}. For the same reason, it is also more sensitive
to uncertainties in hadronic input parametes,
most importantly the inverse moment
$\lambda_B^{-1}$ of the $B$-meson light-cone distribution amplitude.
Altogether, perturbation theory appears to be well behaved and
significant changes in the predictions for branching
fractions and CP asymmetries at this time will be due mainly to changes of
hadronic input parameters such as form factors and light-cone
distribution amplitudes.

\section{Phenomenological implications}
\begin{figure}
\centerline{\psfig{file=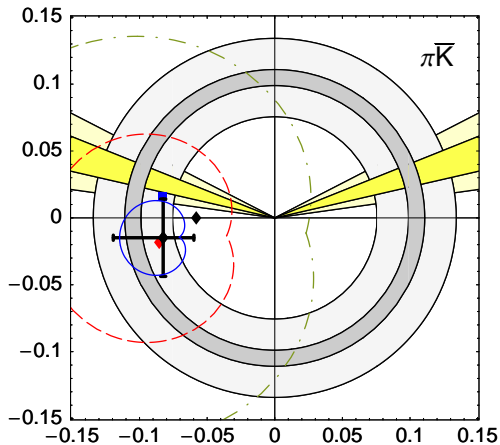,width=2.6in}
}
\caption{Comparing the $PP$ penguin amplitude to data. For an
explanation, see the text. }
\label{fig2}
\end{figure}
Because of the smallness of the corrections compared to uncertainties
due to hadronic input parameters, we do not give updated numbers
for the branching fractions here.
Instead we consider a penguin-to-tree ratio, for which part
of the nonperturbative uncertainties cancel out, but which
nevertheless can be related to experimental data.
Following Ref.~\refcite{BN03}, Fig.~\ref{fig2} shows the ratio
$\hat \alpha_4^c(M_1 M_2)/(\alpha_1(\pi \pi) + \alpha_2(\pi \pi))$
for the pseudoscalar-pseudoscalar ($PP$) final state $M_1 M_2=\pi \bar
K$.
Here $\hat \alpha_4^c(\pi \bar K) = a_4^c(\pi \bar K)
+ r_\chi^{\bar K} a_6^c(\pi \bar K) + \beta_3^c(\pi \bar K)$,
where $r_\chi^{\bar K} a_6^c(\pi \bar K)$ is a numerically large
``charming penguin'' power correction that factorizes
at ${\cal O}(\alpha_s)$
and $\beta_3^c(\pi \bar K)$ models the 
(within QCD factorization) incomputable penguin annihilation
amplitude. The cross shows the theoretical
prediction with errors combined in quadrature (the onion-shaped
regions are various estimates of the annihilation contribution, the
blue one corresponding to the expected magnitude for this power
correction). The grey ring and yellow wedge can be inferred from
data on $\bar B \to \pi \pi$ and $\bar B \to \pi \bar K$
with very little theory input, where the lighter-coloured
areas also including a generous uncertainty on modulus and
phase of $V_{ub}$. The wedge opening to the right is disfavoured by data.
We observe that theory and experiment, which includes $A_{CP}(\bar B^0
\to \pi^+ K^-)$, agree within errors, which is nontrivial.
Some annihilation contribution
is needed, but at the level expected for a power correction. It is
conveivable that a large one-loop spectator-scattering correction
to $a_6^c$ might have a similar impact.

In conclusion, factorization works after inclusion of NLO
spectator-scattering effects. The corrections are small except for the
colour-suppressed electroweak penguin amplitude. The penguin-to-tree
ratios relevant to $\Delta S=1$ decays are consistent with data at the
level of a power correction.

\balance

\section*{Acknowledgements}
The speaker thanks the organizers and his collaborator
Martin Beneke.
This work was supported in part by the DFG
Sonderforschungsbereich/Transregio 9 ``Computergest\"utzte
Theoretische Teilchenphysik''.


\end{document}